\documentclass[final,5p,times,twocolumn]{elsarticle}
\biboptions{square,numbers}

\usepackage{amsmath}
\usepackage{amssymb}
\usepackage{lipsum}
\usepackage{amsthm}
\newtheorem*{Riemann}{Theorem}

\begin{document}

\begin{frontmatter}



\title{Gravity Equivalent to Teleparallelism: Some  Basic Geometrical  Aspects}


\author[first]{Rodrigo Francisco dos Santos}
\ead{santosst1@gmail.com}
\affiliation[first]{organization={Observatório Nacional},
            addressline={R. General José Cristino, 77, São Cristóvão}, 
            city={Rio de Janeiro},
            postcode={20921-400}, 
            state={RJ},
            country={Brazil}}
\author[2nd]{Marilton Rafael da Silva Lemes}
\ead{mariltonrafael@gmail.com}
\affiliation[2nd]{Instituto Federal Fluminense,
            addressline={Estrada Cabo-Frio-Búzios,  s/nº, Baía Formosa}, 
            city={Cabo Frio},
            postcode={28909-971}, 
            state={RJ},
            country={Brazil}}
\author[3rd]{Luis Gustavo de Almeida}
\ead{luis.almeida@ufac.br}
\affiliation[3rd]{organization={Universidade Federal do Acre},
            addressline={BR 364 km 04,  Distrito Industrial}, 
            city={Rio Branco},
            postcode={69920-900}, 
            state={AC},
            country={Brazil}}

\begin{abstract}
We review the book of  Ruben Aldrovandi  and Jose Geraldo Pereira about Teleparallel Gravity. Teleparallel Gravity is an alternative to General Relativity to describe the gravitational interaction.  The difference between General Relativity and Teleparallel Gravity is the fact that  General Relativity associates the curvature to the gravitational interaction, whereas in the Teleparallel Gravity the mathematical object  associated to gravitational field is the torsion, whose field is the so called Weintzenböck connection.  We discuss briefly the advantages of Teleparallel Gravity over General Relativity as well as Weintzenböck connection properties.
\end{abstract}




\end{frontmatter}




\section{Introduction}
\label{introduction}

General Relativity is currently the main theory for describing the gravitational field. 
In General Relativity, the gravitational field is associated with the curvature of space-time and the mathematical object that describes the curvature is the Ricci tensor, which is the contraction of two indices of the Riemann tensor.  This representation comes from Fundamental Theorem of Riemannian Geometry.
\subsection{Fundamental Theorem of Riemannian Geometry}
The Riemann tensor is the mathematical entity that measures the curvature in a Riemannian variety~\citep{will} or pseudo-Riemannian in the case of the varieties used in General Relativity~\citep{sean}.

Following Willmore~\citep{will}, we now write the \emph{Fundamental Theorem of Riemannian Geometry}:
\begin{Riemann}\nonumber
Let $M$ be a pseudo-Riemannian variety. There is a unique connection~$\nabla\in {M}$, such that we can define:
\begin{enumerate}
\item{Torsion is zero;}
\item{Parallel transport preserves the domestic product.}
\end{enumerate}
\end{Riemann}

The theorem allows us to make the definition below:\\
For any 4-vector $v^{\alpha}$ we have, in the absence of torsion,
\begin{equation}\label{riem}
[\nabla_{\mu}, \nabla_{\nu}]v^{\alpha}=R^{\alpha}_{\mu\nu\beta}v^{\beta},
\end{equation}
where  $[\nabla_{\mu}, \nabla_{\nu}]:=\nabla_{\mu}\nabla_{\nu}-\nabla_{\nu}\nabla_{\mu}$ is the covariant derivative switch. That implies the construction of geodesics and the variation of the tangent vector along the geodesic, 
\begin{equation}\label{c1}
\nabla_{\nu}\nabla_{\mu}U_{\beta}=\nabla_{\nu}(\partial_{\mu}U_{\beta})-(\nabla_{\nu}\Gamma^{\alpha}_{\mu\beta})U_{\alpha}-\Gamma^{\alpha}_{\mu\beta}\nabla_{\nu}U_{\alpha}. 
\end{equation}

Swapping the indices $\mu\leftrightarrow\nu$, one have
\begin{equation}\label{c2}
\nabla_{\mu}\nabla_{\nu}U_{\beta}=\nabla_{\mu}(\partial_{\nu}U_{\beta})-(\nabla_{\mu}\Gamma^{\alpha}_{\nu\beta})U_{\alpha}-\Gamma^{\alpha}_{\nu\beta}\nabla_{\mu}U_{\alpha}.
\end{equation}

Subtracting~(\ref{c2}) from~(\ref{c1}), we have
\begin{align}\nonumber
\left[\nabla_{\mu},\nabla_{\nu}\right]U_{\beta}=& \nabla_{\nu}\partial_{\mu}U_{\beta}-\nabla_{\mu}\partial_{\nu}U_{\beta}-\left[\nabla_{\nu}\Gamma^{\alpha}_{\mu\beta}-\nabla_{\mu}\Gamma^{\alpha}_{\nu\beta}\right]U_{\alpha}+\\\label{c1c2}
&+\Gamma^{\alpha}_{\nu\beta}\nabla_{\nu}U_{\alpha}-\Gamma^{\alpha}_{\mu\beta}\nabla_{\nu}U_{\alpha}.
\end{align}

For better clarification, it is convenient to study each term carefully. One can pick the first and second terms of the equation~(\ref{c1c2})
\begin{equation}\label{c11}
\nabla_{\nu}(\partial_{\mu}U_{\beta})=\partial_{\nu}\partial_{\mu}U_{\beta}-\Gamma^{\lambda}_{\nu\mu}\partial_{\lambda}U_{\beta}-\Gamma^{\alpha}_{\mu\beta}\partial_{\nu}U_{\alpha},
\end{equation}
swapping the indices $\mu\leftrightarrow\nu$, we have
\begin{equation}\label{c21}
\nabla_{\mu}\partial_{\nu}U_{\beta}=\partial_{\mu}\partial_{\nu}U_{\beta}-\Gamma^{\lambda}_{\mu\nu}\partial_{\lambda}U_{\beta}-\Gamma^{\alpha}_{\nu\beta}\partial_{\mu}U_{\alpha}.
 \end{equation}

One can see that $\left[\partial_{\nu},\partial_{\mu}\right]U_{\beta}=0$ for flat space. To move the parallel vector in different directions does not make any difference, while the term $\left(\Gamma^{\lambda}_{\nu\mu}-\Gamma^{\lambda}_{\mu\nu}\right)\partial_{\lambda}U_{\beta}$ is zero, because we are dealing with a torsion-free scenario $\left(\Gamma^{\lambda}_{\nu\mu}=\Gamma^{\lambda}_{\mu\nu}\right)$.

The third term is written below
\begin{equation}\label{c12}
\Gamma^{\alpha}_{\mu\beta}\nabla_{\nu}U_{\alpha}=\Gamma^{\alpha}_{\mu\beta}\partial_{\nu}U_{\alpha}-\Gamma^{\alpha}_{\mu\beta}\Gamma^{\kappa}_{\alpha\nu}U_{\kappa},
\end{equation}
and it is easy to see that the term $\Gamma^{\alpha}_{\mu\beta}\partial_{\nu}U_{\alpha}$ from (\ref{c12}) cancels with the corresponding term in the equation (\ref{c11}). Swapping the indices  $\mu\leftrightarrow\nu$, we have the fourth term
  \begin{equation}\label{c22}
\Gamma^{\alpha}_{\nu\beta}\nabla_{\mu}U_{\alpha}=\Gamma^{\alpha}_{\nu\beta}\partial_{\mu}U_{\alpha}-\Gamma^{\alpha}_{\nu\beta}\Gamma^{\kappa}_{\alpha\mu}U_{\kappa}.
 \end{equation}

The same cancellation that occurs between (\ref{c12}) e (\ref{c11}), also occurs between the corresponding terms of (\ref{c21}) and~(\ref{c22}).  Returning to the terms (\ref{c11}), (\ref{c12}), (\ref{c21}) and (\ref{c22}) in (\ref{c1c2}), and  considering that the associated terms~$\partial_{\alpha}U_{\beta}$ are the torsion terms, which by construction are null, we can rewrite (\ref{c1c2}) as
\begin{equation}\label{riemann}
\left[\nabla_{\mu},\nabla_{\nu}\right]U_{\alpha}=\left[\nabla_{\nu}\Gamma^{\beta}_{\mu\alpha}-\nabla_{\mu}\Gamma^{\beta}_{\nu\alpha}+\Gamma^{\kappa}_{\mu\alpha}\Gamma^{\beta}_{\nu\kappa}-\Gamma^{\kappa}_{\nu\alpha}\Gamma^{\beta}_{\mu\kappa}\right]U_{\beta}.
\end{equation}

We finally write the Riemann tensor in terms of the connections
\begin{equation}
R^{\beta}_{\nu\alpha\mu}=\nabla_{\nu}\Gamma^{\beta}_{\mu\alpha}-\nabla_{\mu}\Gamma^{\beta}_{\nu\alpha}+\Gamma^{\kappa}_{\mu\alpha}\Gamma^{\beta}_{\nu\kappa}-\Gamma^{\kappa}_{\nu\alpha}\Gamma^{\beta}_{\mu\kappa}.
\end{equation}

The quantities $\Gamma^{\alpha}_{\beta\kappa}$, are known as connections, representing exactly the curvature correction when we move the parallel vector to itself. Now let's consider the metric tensor $g_{\mu\nu}$, one can apply the derivative to the covariant $g_{\mu\nu}$. And then we have:
\begin{align}\label{gamao}
\nabla_{\alpha}g_{\mu\nu}&=\partial_{\alpha}g_{\mu\nu}+\Gamma^{\beta}_{\mu\nu}g_{\beta\alpha}+\Gamma^{\beta}_{\alpha\mu}g_{\beta\nu}; \\\label{gama2}
\nabla_{\mu}g_{\nu\alpha}&=\partial_{\mu}g_{\nu\alpha}+\Gamma^{\beta}_{\nu\alpha}g_{\beta\mu}+\Gamma^{\beta}_{\mu\nu}g_{\beta\alpha}; \\\label{gama3}
\nabla_{\nu}g_{\alpha\mu}&=\partial_{\nu}g_{\alpha\mu}+\Gamma^{\beta}_{\alpha\mu}g_{\beta\nu}+\Gamma^{\beta}_{\nu\alpha}g_{\beta\mu}.
\end{align}

Adding (\ref{gamao}) and (\ref{gama2}) and subtracting (\ref{gama3}), remembering that in a pseudo-Riemannian variety $\nabla_{\alpha}g_{\mu\nu}=\nabla_{\mu}g_{\nu\alpha}=\nabla_{\nu}g_{\alpha\mu}=0$, we then can write
\begin{equation}
\nabla_{\alpha}g_{\mu\nu}+\nabla_{\mu}g_{\nu\alpha}-\nabla_{\nu}g_{\alpha\mu}=0,
\end{equation}
where we also have
\begin{align}\nonumber
&\partial_{\alpha}g_{\mu\nu}+\partial_{\mu}g_{\beta\nu}-\partial_{\nu}g_{\alpha\mu}+\\\nonumber 
&+\left(\Gamma^{\beta}_{\nu\alpha}-\Gamma^{\beta}_{\alpha\nu}\right)g_{\beta\nu}+2\Gamma^{\beta}_{\mu\nu}g_{\beta\alpha}+\left(\Gamma^{\beta}_{\nu\alpha}-\Gamma^{\beta}_{\alpha\nu}\right)g_{\beta\mu}=0,
\end{align}
 
 If we consider $\Gamma^{\beta}_{\nu\alpha}=\Gamma^{\beta}_{\alpha\nu}$, we are left with the connection in terms of the derivatives of the metric 
\begin{equation}\label{conex}
\Gamma^{\beta}_{\alpha\mu}=\frac{1}{2}g^{\beta\nu}\left(\partial_{\alpha}g_{\mu\nu}+\partial_{\mu}g_{\nu\alpha}-\partial_{\nu}g_{\alpha\mu}\right). 
\end{equation}

In Teleparallel Gravity, the gravitational field is represented by the torsion~\citep{sean}. Both quantities are associated with the parallel transport of a vector $u^{\mu}$
\begin{equation}\label{trc}
\left[\nabla_{\mu}, \nabla_{\nu}\right]u_{\kappa}=R^{\alpha}_{\mu\nu\kappa}u_{\alpha}- T^{\lambda}_{[\mu\nu]}\nabla_{\lambda}u_{\kappa}.
\end{equation}

\section{Building Tetrads}
In this section we follow the book of Aldrovandi and Pereira, see~\citep{Aldro}.

\subsection{Trivial Frames}
We need to discriminate some trivial objects before we can write the Weintzenböck connection.
The metric assigned is
\begin{equation}
\eta_{ab}=diag\left(+1,-1,-1,-1\right).
\end{equation}

Latin indices are related to tangent space,
 greek indices are related to space-time. 
Poincaré Group regulates the transformation for translation to different tangent spaces
\begin{equation}
\mathcal{P}=\mathcal{L}\oslash\mathcal{T},
\end{equation}
wherein $\mathcal{T}$ is translation operator, $\mathcal{L}$ is Lorentzian operator associated to boosts and rotation of group $SO(1,3)$,  
$\{x^{\mu}\}$ are the coordinates in manifold and $\{x^{a}\}$ coordinates in tangent-space.

Whit the gradients $\partial_{\mu}\equiv\frac{\partial}{\partial{x^{\mu}}} $  and $\partial_{a}\equiv\frac{\partial}{\partial{x}^{a}}$, the basis $dx^{\mu}$ and co-vectors  or differentials~$dx^{a}$ become 
\begin{equation}
dx^{\mu}\left(\partial_{\nu}\right)=\delta^{\nu}_{\mu},\quad dx^{a}\left(\partial_{b}\right)=\delta^{a}_{b}. 
\end{equation}
 
This structure of bases allows us to write the tetrads, which are also known as vielbeins
\begin{equation}
e^{a},\quad  e_{a} . 
\end{equation}

One can pick very particular cases of the mentioned ``coordinate'' bases
\begin{equation}\label{tet}
e_{a}=\partial_{a}, \quad e^{a}\equiv{dx}^{a},
\end{equation}
whose name stems from their relation to a coordinate system. Any other set of
four linearly independent fields $e_{a}$ will form another basis, and will have a dual
\begin{equation}\label{t}
e_{a}=e^{\mu}_{a}\partial_{\mu} ,\quad  e^{a}=e^{a}_{\mu}dx^{\mu},
\end{equation}
whose members are such that,
\begin{align}\label{tt}
\partial_{\mu}&=e^{a}_{\mu}e_{a}, dx^{\mu}=e^{\mu}_{a}e^{a}, \\\label{ttt}
e^{a}_{\mu}e^{\nu}_{a}&=\delta^{\nu}_{\mu}, e^{a}_{\mu}e^{\mu}_{b}=\delta^{a}_{b}.
\end{align} 

One can make the structure
\begin{equation}\label{comu}
[e_{a},e_{b}]=f^{c}_{ab}e_{c},
\end{equation}
where $f^{c}_{ab}$ is the coefficient of anholonomy of frame $e_{\alpha}$. 

The dual expression of the commutation relation above is the Cartan structure equation
\begin{equation}
de^{c}=-\frac{1}{2}f^{c}_{ab}e^{a}\wedge{e}^{b}.
\end{equation}

The structure coefficients represent the curls of the basis members
\begin{equation}\label{ano}
f^{c}_{ab}=e^{c}_{\mu}\left[e_{a}\left(e^{\mu}_{b}\right)-e_{b}\left(e^{\mu}_{a}\right)\right]=e^{\mu}_{a}e^{\nu}_{a}\left(\partial_{\nu}e^{c}_{\mu}-\partial_{\mu}e^{c}_{\nu}\right).
\end{equation}

The structure coefficients also regulate a preferred frame, in the case $f^{c}_{ab}=0$. 
This is not a local property, in the sense that it is valid everywhere for frames
belonging to this inertial class. 

\subsection{Non-trivial frames} 
The non-trivial frames or tetrads can be written as~$h_{a}$ and~$h^{a}$. They are defined as linear frames whose coefficient of anholonomy~(\ref{ano}) is related to
both inertia and gravitation. There are differences between trivial  $e_{a}$ and non-trivial frames. 
let's consider a pseudo-Riemannian metric $\textbf{g}=g_{\mu\nu}dx^{\mu}{\otimes}dx^{\nu}$. Just like in the equation~(\ref{tet}), we have the tetrad fields
\begin{equation}\label{tf}
h_{a}=h^{\mu}_{a}\partial_{\mu}, h^{\mu}=h^{a}_{\mu}dx^{\mu}, 
\end{equation}
 wich is a linear basis related to tangent metric~$\textbf{g}$ and to the tangent space metric
 \begin{equation}
{\mathbf{\eta}}=\eta_{ab}dx^{a}\otimes{dx^{b}}.
 \end{equation}

Through the relation
 \begin{equation}
\eta_{ab}=g_{\mu\nu}h^{\mu}_{a}h^{\nu}_{b},
 \end{equation}
the tetrad field (\ref{tf}) can be made as a linear frame. The members of tetrad field $h_{a}$ are pseudo-orthogonal by  the pseudo-Riemannian metric $g_{\mu\nu}$. 
 This construction allows us to say that the field of tetrads $h_{a}$ obeys the same conditions~(\ref{t}), (\ref{tt}) and~(\ref{ttt}) as the fields $e_{a}$, including anoholomology conditions~(\ref{ano}) and~(\ref{comu}). 
 We can now build the Weitzenböck connection~\citep{golo,cara2} in the next section.

 \section{The Weitzenböck connection}
 The  Weitzenböck connections represents the alternative conditions for Riemannian Geometry~(\ref{riem}). In equation (\ref{trc}) we have the $R^{\alpha}_{\mu \nu \kappa}=0$ and $T^{\alpha}_{[\mu \nu]} \neq0$~\citep{golo}. In equation (\ref{trc}) we can define
 \begin{equation}\label{torsão}
T^{\alpha}_{[\mu\nu]}=\Gamma^{\alpha}_{\mu\nu}-\gamma^{\alpha}_{\nu\mu}
 \end{equation}
where we have that the symmetry between inferior indices is violated. That violation is generated by torsion.
 This is a more general geometry compared to the Riemannian geometry. In this sense, we can write the generalized  connection 
 \begin{equation}\label{gconex}
\gamma_{\alpha\mu\nu}=\Gamma_{\alpha\mu\nu}+K_{\alpha\mu\nu},
 \end{equation}
 where $K_{\alpha\mu\nu}$ is the contortion tensor and $\Gamma^{\alpha\mu\nu}$  is the usual connection for $\Gamma_{\alpha\mu\nu}=0$. The Weitzenböck connection $\nabla^{w}$ corresponds to zero curvature and nonzero torsion.  We can assume that there exist $n$ covariant vector fields constant with respect the connection $\nabla^{w}$. 

  We apply the Weitzenböck Connection in tetrads
  \begin{equation}
\nabla^{W}_{i}e^{j}_{c}=\partial_{i}e^{j}_{c}+e^{\kappa}_{c}W^{j}_{ik},
  \end{equation}
  covariant vector fields constant with respect the connection $\nabla^{W}_{i}e^{j}_{c}=0$.  
  Therefore 
  \begin{equation}
W^{i}_{jk}=e^{i}_{a}\partial_{j}E^{a}_{k}.
  \end{equation}
 
  The connection $\nabla^{w}$ is a $\textbf{g}$-metric compatible and also parallel transport of dual base.
  That follows from the fact that $\partial_{i}(e^{j}_{a}E^{a}_{k})=\partial_{i}\delta^{j}_{k}=0$, and we have 
  \begin{equation}
\nabla^{w}_{j}E^{a}=\partial_{j}E^{a}_{k}-E^{a}_{i}W^{i}_{jk}=\partial_{j}E^{a}_{k}-E^{a}_{i}\left(e^{i}_{b}\partial_{j}E^{b}_{k}\right)=0.
  \end{equation}

Moreover, using the standard transformation law for connections shows that the components
of the Weitzenböck connection in the moving frame, $W_{abc}$, vanish. 

\section{Summary and conclusions}
We briefly review some aspects of Connection of Weitzenböck, in special the fact that it is a geometry of zero curvature but nonzero torsion. This is a first approach to a program of investigation in geometric aspects of Gravitation Theory.
This first review is closely related to the first chapter of the reference~\citep{Aldro}. 
It corresponds to the first steps of a project to teach differential geometry and gravity, one of the central tools of which is the Aether Tenebris channel on YouTube. Where the first author presents reviews of books on Gravitation, General Relativity and Hydrodynamics, as well as other topics related to astrophysics and mathematics.

\section*{Acknowledgements}
Thanks to professor Eraldo Almeida Lima Junior from Universidade Federal da Paraíba for the encouragement and discussions.

\bibliography{ReferencesTERG}





.

%
%
%
%
%
%

\end{document}